\documentclass[conference]{IEEEtran}
\IEEEoverridecommandlockouts
\usepackage{cite}
\usepackage{amsmath,amssymb,amsfonts}
\usepackage{algorithmic}
\usepackage{graphicx}
\usepackage{textcomp}
\usepackage{ulem}
\usepackage{xcolor}

\renewcommand{\figurename}{Figure}
\usepackage[font=footnotesize,labelsep=period]{caption} 

\usepackage{subcaption}
\def\BibTeX{{\rm B\kern-.05em{\sc i\kern-.025em b}\kern-.08em
    T\kern-.1667em\lower.7ex\hbox{E}\kern-.125emX}}

\newcommand{\imag}[0]{\mathrm{i}\mkern1mu}

\usepackage{hyperref}

\begin{document}

\title{The Risk of Cascading Failures in Electrical Grids Triggered by Extreme Weather Events \\

\thanks{This project has received funding from the CoNDyNet2 project under grant no. 03EF3055F}
}

\author{\IEEEauthorblockN{Julian M. St{\"u}rmer}
\IEEEauthorblockA{\textit{(1) Complexity Science department,}\\
\textit{Potsdam Institute for}\\
\textit{Climate Impact Research}\\
Potsdam, Germany\\
\textit{(2) Institut f{\"u}r Theoretische Physik, TU Berlin}\\
Berlin, Germany \\
email: Julian.Stuermer@pik-potsdam.de}
\and
\IEEEauthorblockN{Anton Plietzsch}
\IEEEauthorblockA{\textit{(1) Complexity Science department,}\\
\textit{Potsdam Institute for}\\
\textit{Climate Impact Research} \\
Potsdam, Germany\\
\textit{(2) Humboldt-Universit{\"a}t zu Berlin}\\
Berlin, Germany \\
email: plietzsch@pik-potsdam.de}
\and
\IEEEauthorblockN{Mehrnaz Anvari}
\IEEEauthorblockA{\textit{Complexity Science department,}\\
\textit{Potsdam Institute for}\\
\textit{Climate Impact Research}\\
Potsdam, Germany \\
email: anvari@pik-potsdam.de}
}

\maketitle

\begin{abstract}
One of the serious threats related to climate change is an increase in the number and severity of extreme weather events. A prominent example are hurricanes, which result from rising coastal temperatures.
Such extreme weather events can cause extensive damages in infrastructure systems and, potentially, destroy components in electricity transmission networks, which in turn can lead to major blackouts.
In our recent work, we study the risk of hurricane-induced cascading failures in power systems of the U.S. East Coast using historical wind field data sets. For this purpose, we model the destruction of overhead transmission lines during hurricanes, where each failing line causes a rerouting of power flow in the system. New power
flows can overload additional lines, which are then automatically deactivated and thereby cause another rerouting of power flow and so on. Ultimately, a cascade of failures can unfold that can black out large parts of the power system.
\end{abstract}


\section{Introduction}

Recent climate studies demonstrate that human-induced climate change not only causes a rapid increase in global temperature but also leads to more frequent and severe extreme weather events, such as floods, heavy rainfall, winter storms and hurricanes \cite{b1}.
These events threaten the stability of social facilities and services, as well as social network infrastructures, such as public health care, transportation, telecommunication, and electrical grids. 
Due to the extensive economic and social consequences that accompany disruptions in these systems, fostering their resilience and thereby mitigating the impact of extreme weather events represents an important challenge for governments and societies.

It is worth mentioning that, due to the intertwined nature of social infrastructure systems, a failure in one system can easily spread into other systems. This especially applies to failures in electrical grids, since they can lead to major blackouts that impair the access to food, transportation, medical treatment and so on.
In recent years, reports demonstrated that the increase in extreme weather events puts electrical grid components at greater risk of failure in several parts of the world.
As an example, the authors in \cite{b2} discuss how the probability of damages in the British transmission network increases with more frequent winter storms. In addition, \cite{b3} shows how transmission tower damages in Australia can be attributed to localized downbursts, that become more often.

Furthermore, recent data recorded in the Atlantic Ocean \cite{b4,b5} show an increase in coastal temperatures that promote more frequent and intense hurricanes in coastal states of the U.S. like Texas and Louisiana. For instance, hurricane Laura led to major outages in Louisiana and other states in the days following the August 27th, 2020 \cite{b6}.
Since the restoration of destroyed components in electrical grids can be very costly for power transmission authorities (typically ranging from millions to billions of euros), identifying vulnerable components and improving the resilience represent a promising way of avoiding high costs and extensive outages.

In our ongoing research, we apply a quasi-static model to investigate the risk of hurricane-induced blackouts in a synthetic electrical grid for Texas. In Sec.~ \ref{sec_pg_modeling}, we discuss how power flows can be analyzed in the electrical grid. Next, we establish a probabilistic approach in Sec.~\ref{sec_wind_induced_damage} to model wind-induced failures of overhead transmission lines occurring in the course of a traversing hurricane. Our work incorporates the analysis of power flows that change after each failure and the simulation of cascading failures by deactivating lines and transformers as soon as they become overloaded. Preliminary results of blackouts will be presented in Sec.~\ref{sec_cascades}. 

\section{Electrical Grid Modeling}
\label{sec_pg_modeling}

To assess the impact of extreme weather events in realistic scenarios, an elementary question arising is how and in what detail the electrical grid should be modeled. Regarding to the power system research, an electrical grid is a network with $N$ nodes, representing generators and loads, and $E$ edges, representing transmission lines and transformers. We will also refer to nodes as buses and to edges as branches throughout this paper.

In the context of cascading failures, both steady-state models, as well as dynamic models have been used for calculating the power flow on branches and it is known that the final outcome of cascades can vary between different models \cite{schaefer_dynamics}. It is therefore not certain how cascading failures should be modeled in general in order to obtain the most realistic and plausible results. Furthermore, the applicability of different power flow models also depends on the availability of grid parameters, such as inertia and damping coefficients in the dynamic swing equation or reactive power injections in the static AC power flow model. Data sets that provide these parameters together with a realistic power grid topology are generally rare because of confidentiality.

The application presented throughout this paper uses a sophisticated synthetic electrical grid data set generated by Birchfield et al. on the footprint of Texas \cite{birchfield_grid_structural}. 
The data set is publicly available under the name ``ACTIVSg2000'' in a test case repository hosted by the Texas A\&M University \cite{test_case_repository}. It encompasses $2000$ buses with geographic locations and an extensive set of parameters, some of which will be introduced in Sec.~\ref{subsec_ac_power_flow}.

The following sections will introduce the AC power flow model and its DC approximation, both of which allow to calculate power flows in the regarded electrical grid of Texas. Using these models to determine the steady states of the power grid enables a quasi-static description of cascading failures that will be discussed in Sec.~\ref{sec_cascades}. 

\subsection{AC Power Flow}
\label{subsec_ac_power_flow}


The AC power flow equations representing a set of $2N$ nonlinear algebraic equations are solved to obtain the voltage magnitudes $|V_i|$ and voltage angles $|\theta_i|$ of all buses $i \in \{1,\dots,N\}$. These voltage variables characterize the steady state of an AC power grid and enable the calculation of all power flows. The AC power flow equations are given by
\begin{align}
    \label{eq_ac_active}
    P_i &= \sum_{j=1}^N |V_i||V_j| [G_{ij} \cos(\theta_i - \theta_j) + B_{ij} \sin(\theta_i - \theta_j)] \, ,\\
    \label{eq_ac_reactive}
    Q_i &= \sum_{j=1}^N |V_i||V_j| [G_{ij} \sin(\theta_i - \theta_j) - B_{ij} \cos(\theta_i - \theta_j)] \, ,
\end{align}
where $P_i \in \mathbb{R}$ and $Q_i \in \mathbb{R}$ represent the active and reactive power injections at each bus. The coefficients $G_{ij} \in \mathbb{R}$ and $B_{ij} \in \mathbb{R}$ are, respectively, the real and imaginary part of the nodal admittance matrix $Y_{ij} = G_{ij} + \imag B_{ij} \in \mathbb{C}$ and incorporate branch conductances and susceptances as well as additional shunt contributions. All of these parameters are provided in the ``ACTIVSg2000'' data set and their values are generally given in the per-unit system (p.u.).
In order to solve the AC power flow equations \eqref{eq_ac_active} and \eqref{eq_ac_reactive}, we draw upon the Julia package PowerModels.jl \cite{powermodelsjl}.



\subsection{DC Approximation}

\begin{figure}[tbp] 
\centerline{\includegraphics[width=\columnwidth]{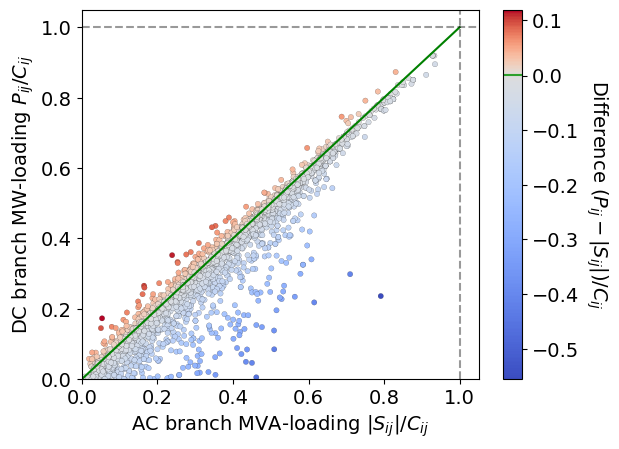}}
\caption{Loading of branches in the DC model versus respective loading in the AC model. Individual data points are colored according to the difference in loading. The diagonal green line indicates the perfect match.}

\label{fig_ac_dc_comparison}
\end{figure}

The DC approximation represents a linearized version of the AC model introduced in Sec.~\ref{subsec_ac_power_flow}. Following three key assumptions have to be made for DC power flow: 
(i) Neglect ohmic losses resulting from resistances, i.e. $G_{ij} = 0\,\mathrm{p.u.}$ for all branches in the network;
(ii) 
Fix the voltage magnitudes to $|V_i| = 1\,\mathrm{p.u.}$ for all buses $i \in \{1,\dots,N\}$. This reduces the number of unknown state variables by $N$ and one can therefore drop the $N$ reactive power equations shown in \eqref{eq_ac_reactive};
(iii) Consider small voltage angle differences $|\theta_i - \theta_j|$ along all branches.
This allows us to linearize the sine function in \eqref{eq_ac_active} and, ultimately, leads to the $N$ DC power flow equations.
\begin{align}
    \label{eq_dc_model}
    P_i = \sum_{j=1}^N B_{ij} (\theta_i - \theta_j) \, .
\end{align}
These equations are also solved for the considered Texas grid using PowerModels.jl. Due to the strong approximations made in the DC model, the resulting power flows tend to differ from the AC solution.
A comparison of both solutions is shown in Fig.~\ref{fig_ac_dc_comparison}, where we plot the loading of branches in the DC model versus their corresponding loading in the AC model.
Since reactive power also contributes to the loading of branches in the AC model, various branches have to transport more power. In the context of cascading failures driven by overloaded branches, this can suggest substantially different cascades in both models.
Nevertheless, the DC model provides better solvability and is solved much faster than the AC model making it a popular choice for computing cascading failures. 
Motivated by Fig.~\ref{fig_ac_dc_comparison}, we are investigating the difference between AC and DC cascades in ongoing work. Sec.~\ref{sec_cascades} presents preliminary results using the DC model.


\section{Wind-Induced Damage}
\label{sec_wind_induced_damage}
To demonstrate the impact of extreme weather events on electrical grids, we here discuss the modeling of wind-induced damages. In principle, as wind speeds exceed the design wind speeds of structures in an electrical grid, damages can be expected to occur in the course of an extreme event. However, it is never certain in advance which components will indeed fail. Hence, probabilistic approaches modeling the fragility of structures with regard to weather-induced failures have been developed throughout the years.

For the hurricane scenarios that we study,
we adopt a probabilistic description of the fragility of overhead transmission lines first introduced by Winkler et al. in \cite{winkler_hurricane_events}. The method is based on the standard wind force design equation defined by the American Society of Civil Engineers (ASCE) in \cite{asce_report} that allows to calculate the wind force $F_\mathrm{wind}$ acting on
transmission line segments. Line segments represent pairs of transmission towers with conductor wires spanned between them. In reality, all overhead transmission lines in the synthetic electrical grid of Texas consist of a large number of line segments. We therefore divide all overhead transmission lines into $N_\mathrm{seg}$ individual segments using the average distance of $161\,\mathrm{m}$ between transmission towers \cite{watson_hurricane_conditions}. Motivated by \cite{winkler_hurricane_events}, we proceed by assigning failure probabilities to individual line segments $k$ according to
\begin{align}
    \label{eq_failure_probability}
    p_k(v, l) = \min \left(\gamma\, \frac{F_{\text{wind},k}(v, l)}{F_{\text{brk},k}},\, 1\right)\, ,
\end{align}
where $v$ is the local $3$-second gust wind speed assumed to act perpendicular to the line direction, $l$ is the obtained wire span length of roughly $161\,\mathrm{m}$ and $F_{\mathrm{brk},k}$ is the maximum perpendicular force that the wire can endure, which is chosen according to the grid data set.
The coefficient $\gamma$ is a scaling parameter proposed in \cite{winkler_hurricane_events}, that can be used to match the average number of wind-induced line failures to historical data. The impact of the scaling parameter will be discussed in the last paragraph of this section.
According to the wind force $F_\mathrm{wind}$ defined in \cite{asce_report}, the failure probabilities in \eqref{eq_failure_probability} increase with the square of the wind speed $p_k \propto v^2$.


Given the wind speed data, $v$, for a specific hurricane, we can generate uniform random numbers in $[0,1]$ for each line segment and time step and compare them to the respective probabilities \eqref{eq_failure_probability}.
Once a random number falls below the failure probability, we remove the corresponding transmission line containing the destroyed segment.
Ultimately, each scenario of a hurricane traversing the grid represents a chronological sequence of wind-induced line failures used to compute cascading failures that may be triggered. As our model is probabilistic, we can perform Monte Carlo simulations to assess the expected damage of the grid for a specific wind data set.
\section{Cascading Failures}
\label{sec_cascades}

\begin{figure*}[htbp]
\centerline{\includegraphics[width=.8\textwidth]{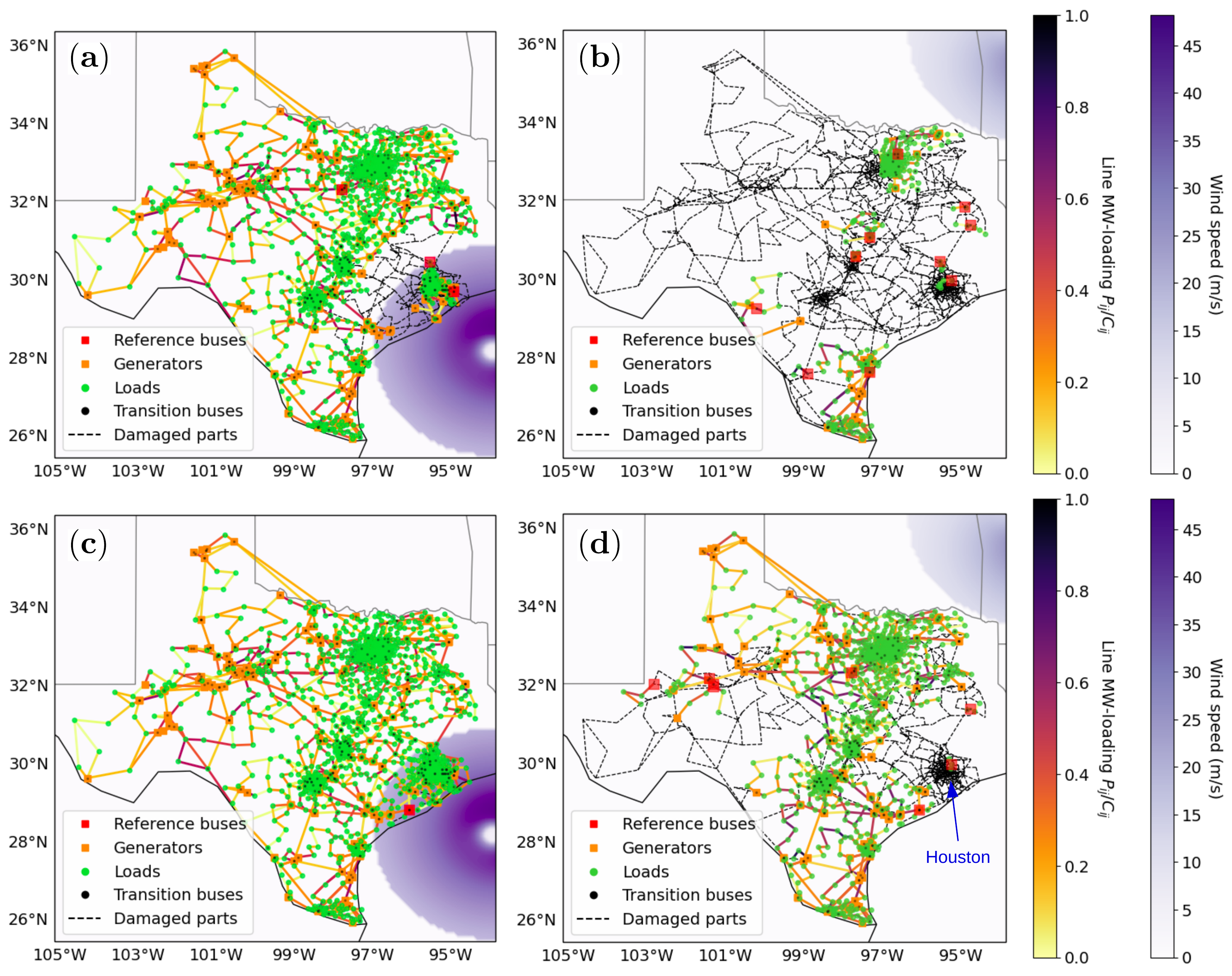}}
\caption{
Two different cascading failure scenarios triggered by hurricane Ike. (a) and (b) show two time steps in a worst case scenario. (c) and (d) show the same time steps in a typical scenario. Houston which is extensively damaged in both worst and typical scenario has been annotated in (d).
}
\label{fig_dc_cascades}
\end{figure*}

In this section, we explore cascading failures that unfold during specific hurricane scenarios described in Sec.~\ref{sec_wind_induced_damage}. For this purpose, the power flow models introduced in Sec.~\ref{sec_pg_modeling} are used to recalculate power flows every time transmission lines are destroyed by the traversing hurricane. We also deactivate overloaded branches that may arise in new power flow solutions and repeat this process until no more overloaded branches persist. Afterwards, our simulations advance to the next wind-induced line failure and so on. This approach can be justified by the fact that cascading failures driven by overloads typically happen on the time scale of seconds while significant changes in the wind speeds typically occur in the course of minutes to hours.
In order to prevent infeasible power flow problems during our simulations, we implement additional control loops that enclose the power flow calculation. This is especially important in the AC model, whenever reactive power is no longer generated in the direct neighborhood of loads. Since refining these loops is still work in progress, we continue by focusing on the simpler DC model. The latter merely requires a single outer loop restoring active power balance after generators and/or loads were disconnected from the grid or the system split into separated islands. Here, active power balance means that the generators cover the total demand in the grid. We have therefore established an algorithm that restores balance motivated by primary frequency control in real systems or deactivates connected components, if balance cannot be restored within the capacity of generators.

Fig.~\ref{fig_dc_cascades} shows two time steps in two scenarios of hurricane Ike that passed over the eastern part of Texas in 2008. The wind data was calculated using the CLIMADA Python implementation \cite{climada_python} and the IBTrACS archive \cite{ibtracs_archive}. Both scenarios were generated using a scaling factor of $\gamma = 0.005$ corresponding to an average number of $189$ wind-induced failures. Fig.~\ref{fig_dc_cascades}(a) and (b) show a worst case scenario that was found by running $10^4$ simulations. The cascading failures triggered in this scenario lead to a final loss of $\sim 81\%$ of the initial active power supply in the grid. In contrast, in the scenario shown in Fig.~\ref{fig_dc_cascades}(c) and (d) $\sim 33\%$ of the initial supply is lost, which seems to be the most probable outcome judging from our $10^4$ runs. It is worth mentioning that, even though hurricane Ike merely hits the eastern part of the grid, the majority of scenarios induce failures in the western part of the grid due to the spreading of overloads. Fig.~\ref{fig_dc_cascades} shows that the outcome of different scenarios can vary drastically and we are investigating what can trigger worst case scenarios like the one discussed. 
\section{Conclusions}
In this paper, we demonstrated that overhead transmission lines destroyed by hurricane Ike trigger cascading failures in a synthetic electrical grid for Texas (see Fig.~\ref{fig_dc_cascades}). As outlined in Sec.~\ref{sec_cascades}, we combine a Monte Carlo-like method modeling wind-induced failures (see Sec.~\ref{sec_wind_induced_damage}) with a DC power flow model to assess the final power outage in different hurricane scenarios.

As an example, we presented a worst case scenario of hurricane Ike in Fig.~\ref{fig_dc_cascades}, in which more than $80\%$ of the initial power supply is lost, together with a most probable scenario, in which around $30\%$ is lost.
In order to prevent the DC power flow problem from becoming infeasible as the electrical grid is destroyed, we applied a control loop enclosing the power flow calculations that restores active power balance motivated by primary frequency control.

Apart from the DC model, we also touched on the more accurate AC model that requires rather sophisticated control strategies to remain feasible during cascading failures. We show in Fig.~\ref{fig_ac_dc_comparison} that AC and DC power flows already differ considerably in the stable initial state of the Texas grid. Motivated by this fact we plan to study different hurricane scenarios using the AC model and compare the final outages to corresponding DC simulations.
Here, we only considered an electrical grid of Texas and wind data belonging to hurricane Ike as a case study. Nevertheless, our approach can be applied to electrical grids in other geographical regions, such as South Carolina or Louisiana, and to other hurricanes as well.


\clearpage


\end{document}